\documentclass[conference,a4paper]{IEEEtran}

\def\BibTeX{{\rm B\kern-.05em{\sc i\kern-.025em b}\kern-.08em
    T\kern-.1667em\lower.7ex\hbox{E}\kern-.125emX}}
\IEEEsettopmargin{t}{54pt}
\pdfpageattr{/Group << /S /Transparency /I true /CS /DeviceRGB >>}

\usepackage[utf8]{inputenc}
\usepackage[T1]{fontenc}
\usepackage{textcomp}
\usepackage{microtype}
\usepackage{calc}
\usepackage[normalem]{ulem}
\usepackage{verbatim}
\usepackage{float}

\usepackage{lipsum}

\usepackage[range-phrase=--,per-mode=symbol-or-fraction,range-units=single,list-units=single,detect-all,forbid-literal-units=true]{siunitx}
\AtBeginDocument{
\DeclareSIUnit\bit{bit}
\DeclareSIUnit\byte{Byte}
\DeclareSIUnit\mbps{\mega\bit\per\second}
\DeclareSIUnit\kmh{\kilo\meter\per\hour}
\DeclareSIUnit\mw{\milli\watt}
\DeclareSIUnit\decibelm{dBm}
\DeclareSIUnit\vehicle{veh}
}

\usepackage{silence}
\usepackage{csquotes}
\usepackage[backend=biber,style=ieee,doi=false,isbn=false,mincitenames=1,maxcitenames=2]{biblatex}
\DeclareFieldFormat{sentencecase}{#1} 
\DeclareFieldFormat{titlecase}{#1} 
\usepackage{xpatch}
\xpatchbibmacro{textcite}{\addspace}{\addnbspace}{}{}
\xpatchbibmacro{Textcite}{\addspace}{\addnbspace}{}{}
\DefineBibliographyStrings{english}{
	andothers = et~al\adddot\addspace
}

\DeclareSourcemap{
	\maps[datatype=bibtex]{
		\map[overwrite=true]{
			\step[fieldsource=school,fieldtarget=institution]
		}
	}
}

\usepackage{amsmath}

\usepackage{amssymb}
\usepackage{amsfonts}
\usepackage{amsthm}

\usepackage{booktabs}
\usepackage{url}

\usepackage[inline]{enumitem}

\usepackage[caption=false,font=footnotesize]{subfig}
\usepackage[pdftex]{graphicx}
\DeclareGraphicsExtensions{.pdf,.png,.jpg}
\usepackage{tikz}
\usetikzlibrary{arrows}
\usetikzlibrary{calc}
\usetikzlibrary{chains}
\usetikzlibrary{scopes}

\usepackage[american]{babel}
\usepackage{hyphenat}

\usepackage{algorithmic}
\usepackage{algorithm}

\usepackage[capitalize,noabbrev,nameinlink]{cleveref}

\usepackage[nodebug]{flushend}

\RequirePackage{xstring}
\RequirePackage{xparse}
\RequirePackage[]{acro}
\makeatletter
\@ifpackagelater{acro}{2019/02/17}{
	\NewDocumentCommand\acrodef{mO{#1}mG{}}{\DeclareAcronym{#1}{short={#2}, long={#3}, foreign-plural={}, #4}}
}{
	\NewDocumentCommand\acrodef{mO{#1}mG{}}{\DeclareAcronym{#1}{short={#2}, long={#3}, #4}}
}
\makeatother

\acrodef{AI}{Artificial Intelligence}
\acrodef{AGV}{Automated Guided Vehicle}
\acrodef{AP}{Access Point}
\acrodef{BER}{Bit Error Rate}
\acrodef{BPSK}{Binary Phase-Shift Keying}
\acrodef{CNC}{Central Network Configuration}
\acrodef{DB}{Database}
\acrodef{FTM}{Fine Timing Measurement}
\acrodef{gPTP}{generic Precision Time Protocol}
\acrodef{IoT}{Internet of Things}
\acrodef{IWN}{Industrial Wireless Network}
\acrodef{KPI}{Key Performance Indicator}
\acrodef{MCS}{Modulation and Coding Scheme}
\acrodef{MLO}{Multi-Link Operation}
\acrodef{OFDMA}{Orthogonal Frequency-Division Multiple Access}
\acrodef{PTP}{Precision Time Protocol}
\acrodef{QoS}{Quality of Service}
\acrodef{RT}{Ray-Tracing}
\acrodef{TF}{Trigger Frame}
\acrodef{TMS}{Theater Management System}
\acrodef{TSN}{Time-Sensitive Networking}
\acrodef{TWT}{Target Wake Time}
\acrodef{UC}{Use Case}
\acrodef{UI}{User Interface}
\acrodef{WLAN}{Wireless Local Area Network}
\acrodef{XR}{eXtended Reality}
\acrodef{ML}{Machine Learning}
\acrodef{OM}{Operation and Maintenance}
\acrodef{NLOS}{Non-Line-of-Sight}
\acrodef{LOS}{Line-of-Sight}
\acrodef{mmWave}{Millimeter-Wave}
\acrodef{UAV}{Unmaned Aerial Vehicle}

\acrodef{BTS}{Base Station}
\acrodef{DT}{Digital Twin}
\acrodef{DTN}{Digital Twin Network}
\acrodef{RIS}{Reconfigurable Intelligent Surfaces}
\acrodef{RRC}{Radio Resource Connection}
\acrodef{RSSI}{Radio Signal Strength Indicator}
\acrodef{RTWP}{Received Total Wideband Power}
\acrodef{SINR}{Signal to Interference Noise Ratio}
\acrodef{RF}{Radio Frequency}
\acrodef{EMF}{Electromagnetic Field}
\acrodef{RSRP}{Received Signal Received Power}
\acrodef{OFDM}{Orthogonal Frequency Division Multiplexing}
\acrodef{UT}{User Terminal}
\acrodef{3GPP}{3rd Generation Partnership Project}
\acrodef{MNO}{Mobile Network Operator}
\acrodef{OSS}{Operation Support System}
\acrodef{CTTC}{Centre Tecnol\`ogic de Telecomunicacions de Catalunya}
\acrodef{MSE}{Mean Squared Error}
\acrodef{DoS}{Denial of Service}
\acrodef{mMIMO}{Massive MIMO}
\acrodef{UMa}{Urban-Macrocell}
\acrodef{UMi}{Urban-Microcell}

\usepackage{todonotes}

\NewDocumentCommand\IEEE{ s m >{\SplitArgument{4}{/}}d[] }{%
	\IfBooleanTF{#1}{}{IEEE\,}
	\nolinebreak[2]
	#2%
	\IfNoValueTF{#3}{%
	}{%
		\sommerIEEELettersSlashed#3%
	}%
}
\newcommand{\sommerIEEELettersSlashed}[5]{%
	\IfNoValueTF{#2}{%
	}{%
		\nolinebreak[3]
	}%
	#1%
	\IfNoValueTF{#2}{}{/#2}%
	\IfNoValueTF{#3}{}{/#3}%
	\IfNoValueTF{#4}{}{/#4}%
	\IfNoValueTF{#5}{}{/#5}%
}

\usepackage{pifont}
\usepackage{multirow}

\begin{document}

\title{Digital Twin for Advanced Network Planning: Tackling Interference}

\author{%
\IEEEauthorblockN{%
    Juan C. Estrada-Jim\'enez\IEEEauthorrefmark{1},~%
    Valdemar R. Farr\'e-Guijarro\IEEEauthorrefmark{2},~%
    Diana C. Alvarez-Paredes\IEEEauthorrefmark{3},~%
    Marie-Laure Watrinet\IEEEauthorrefmark{1}%
}%
\IEEEauthorblockA{%
    \IEEEauthorrefmark{1}IT for Innovative Services Department, Luxembourg Institute of Science and Technology, Luxembourg%
    \\
    \IEEEauthorrefmark{2}Dpto. Electr\'onica, Telecomunicaciones y Redes de Informaci\'on, Escuela Polit\'ecnica Nacional, Ecuador%
    \\
    \IEEEauthorrefmark{3}STARION Group, Luxembourg%
}%
\IEEEauthorblockA{%
    \small\texttt{%
        \{juan.estrada-jimenez, marie-laure.watrinet\}@list.lu,
    }%
}%
\IEEEauthorblockA{%
    \small\texttt{%
        valdemar.farre@epn.edu.ec, 
        d.alvarez@stariongroup.eu%
    }%
}%
}
\maketitle

\begin{abstract}\nohyphens{%
Operational data in next-generation networks offers a valuable resource for Mobile Network Operators to autonomously manage their systems and predict potential network issues. Machine Learning and Digital Twin can be applied to gain important insights for intelligent decision-making. This paper proposes a framework for Radio Frequency planning and failure detection using Digital Twin reducing the level of manual intervention. In this study, we propose a methodology for analyzing Radio Frequency issues as external interference employing clustering techniques in operational networks, and later incorporating this in the planning process. Simulation results demonstrate that the architecture proposed can improve planning operations through a data-aided anomaly detection strategy.
}\end{abstract}


\acresetall
\IEEEpeerreviewmaketitle

%


\section{Introduction}
\label{sec:intro}
The deployment of 5G and beyond networks presents new challenges for \acp{MNO}, due to its complex nature and evolving requirements. As new features are introduced, automation becomes even more crucial for efficient networks. Moreover, the substantial volume of data generated by 5G networks necessitates intelligent solutions capable of handling the dynamic network performance to detect abnormal patterns. To tackle this, novel data processing techniques enable the utilization of \ac{ML} to optimize deployed 5G mobile networks \cite{6G_ML}. 

In the coming 6G, it is essential to focus on the data patterns provided by the operational data. By understanding the context surrounding the data, including location, user behavior, environmental, and network conditions, mobile networks can enhance performance and improve user experiences \cite{con_2020}. This context-driven approach empowers networks to dynamically adapt and take informed real-time decisions, leading to optimized resource allocation, efficient routing, intelligent network management, and personalized service provisioning. 
\ac{ML} techniques can also enhance decision-making and monitoring in mobile networks to identify and diagnose network faults. Both 5G and 6G networks face challenges related to \ac{RF} interference and network faults. With the increasing demand for higher data rates and connectivity, wireless communication systems become more complex. In 5G networks, \ac{RF} interference can disrupt the quality of service, impacting network performance and degrading user experiences. Detecting and mitigating \ac{RF} interference is crucial to ensure reliable and uninterrupted communications \cite{int_2023}. 

The concept of \ac{DT} will revolutionize mobile networks. In the case of network planning, the \ac{DTN} can provide valuable information for design and diagnosis in 6G networks. This consists of a new approach that creates a replica of mobile networks that allows a continuous testing scenario \cite{dt_net1}. The \ac{DT} uses interfaces to interconnect between physical and virtual representations.

Effectively addressing \ac{RF} interference and network faults in both 5G and 6G networks necessitates the development of sophisticated monitoring and diagnostic tools. These tools should leverage \ac{ML} and data analytics techniques to identify and classify interference, predict potential network faults, and enable proactive measures for optimizing network performance. 

In this paper, we propose a methodology for analyzing \ac{RF} abnormalities from \ac{KPI}, leading to the detection of interfering sources. The proposed methodology employs clustering techniques to classify \ac{KPI} time series based on the \ac{RF} contribution of interfering sources and their correlations with neighboring cells. The framework considers the cells' \ac{RSSI} to recognize patterns that can later be used to predict the presence of a possible \ac{RF} error. Later, data-aided planning is presented using a \ac{DT} approximation. Additionally, we carry out a \ac{RF} planning process aided by a \ac{DT} framework and we provide a comprehensive guide on interference detection using interference \acp{KPI}. The rest of the paper is organized as follows. Section II shows an overview of the radio channel challenges. A brief introduction to propagation models and interference sources is presented. In section III, a \ac{DT} architecture for network planning is proposed and described. Section IV describes a network planning framework considering a real use case, coverage area planning, the management operation, and an extension to a 6G scenario. Finally, the results and conclusions are presented in sections V and VI, respectively.  

%

\section{Overview of Radio Channel Challenges}
\label{subsec:rad_cha}
According to the principle in \cite{emp6g_dt} for using the DT technique for assistance and studies of propagation in Terahertz and the operation of RISs in 6G communication systems, the following aspects are considered:

\subsection{\ac{LOS}/\ac{NLOS}}
\label{subsec:los_nlos}
In the evolving landscape of 6G networks, the difference between \ac{LOS} and \ac{NLOS} communications becomes more important. In the case where there is \ac{LOS} between the transmitter and the receiver, can directly see each other. The radio waves in new higher spectrum work better due to their degrading pathloss, therefore, LOS becomes more important to offer good performances such as low latency and higher capacity. By other side, in \ac{NLOS},  where there is no direct path between the transmitter and receiver, there are a lot of challenges due to signal attenuation and multipath propagation. Several solutions have been proposed recently to face these problems, such as \ac{RIS}, positioning reflective elements in strategic locations to redirect the electromagnetic waves \cite{cha_mod_2021} to the desired destination point.

\begin{figure}
\centering
\includegraphics[width=0.35\textwidth]{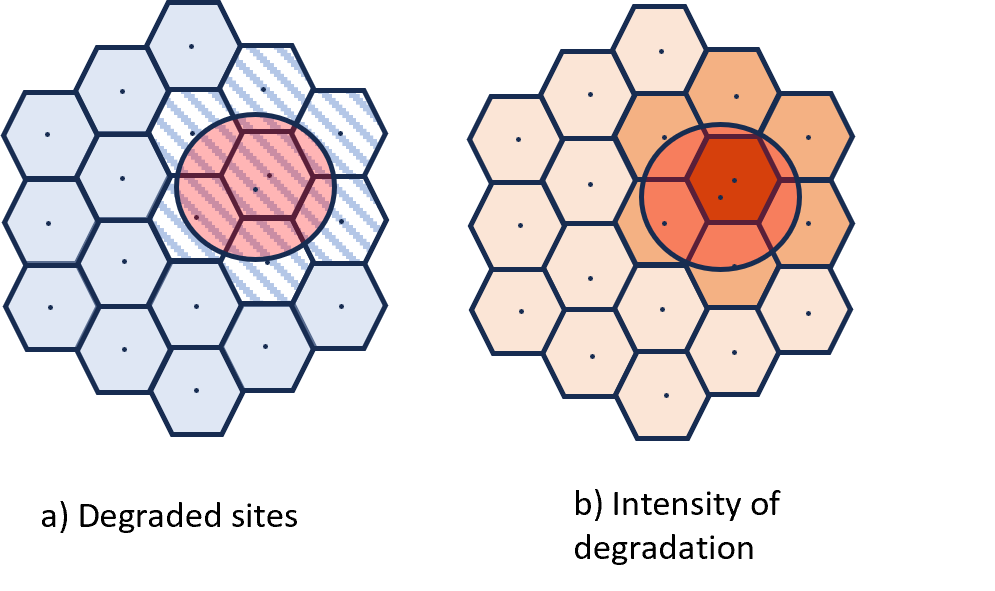}
\vspace{-3mm}
\caption{External interference in RF environments.}\label{img:ext_int}
\vspace{-2mm}
\end{figure}

\vspace{-1mm}
\subsection{6G propagation models}
\label{subsec:pro_mod}

Propagation models are used to evaluate how electromagnetic waves travel through the wireless medium. These models consider several factors, such as the properties of the terrain, obstacles, antennas, atmospheric conditions and others. Additionally, these models predict signal strength, pathloss fading and crucial effects  for designing and optimizing wireless communication systems.
The last communications generations have been updating their propagation models, considering new characteristics such as new frequencies with their corresponding losses, new antenna technologies, and other aspects. Standardized \ac{3GPP} propagation models have been introduced in \cite{3GPPTR38901}. Other more realistic models have been developed using ray-tracing techniques \cite{seidel1992ray} and a low-complex version in \cite{sandouno2023novel}. 

New emerging technologies, such as \ac{mmWave} and terahertz, have been investigated as part of the 5G and 6G efforts \cite{HUANG2023106}. These channel frequencies show great advantages in terms of data rate, bandwidth and directivity, but with significant path losses. By other side, optical wireless communications, which operates in infrared and visible light spectrum, offers distinct channel characteristics compared to traditional frequency bands with more complex scattering properties \cite{VLC}. This new frequency spectrum has unique channel properties for LOS and NLOS but it does not experience multipath fading and Doppler effects. 
A propagation model for 6G would need to address the emerging channel's unique characteristics. These models will need to face more complex propagation phenomena such as scattering, environmental absorption and other effects. The new propagation models for 6G will be able to adapt to new planning techniques and antenna technologies, such as Cell-Free \ac{mMIMO}, Holographic MIMO \cite{holo_antena}, \ac{RIS}, and Fluid antenna systems \cite{fuid_antenna}.

The same way as in previous generations, the new 6G propagation models will have to consider current propagation models and incorporate new features coming from the new emergency frequency bands.

\vspace{-1mm}
\subsection{External interference in RF environments}
\label{subsec:ext_int}

\begin{figure}
\centering
\includegraphics[width=0.48\textwidth]{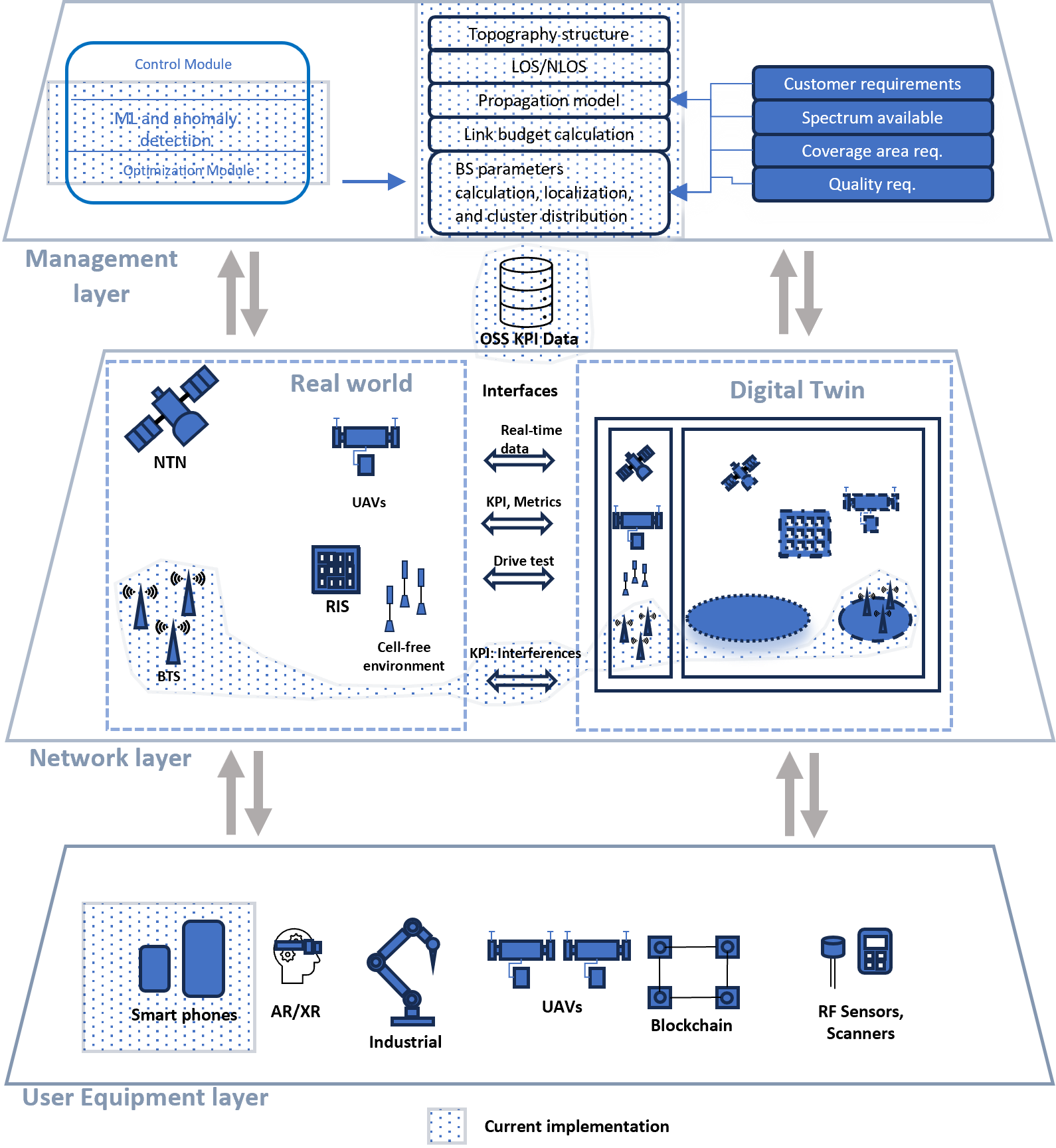}
\vspace{-2mm}
\caption{\ac{DT}-Based Architecture for Planning Next-G Mobile Networks.}\label{img:architecture}
\vspace{-3mm}
\end{figure}

Several sources of interference can be present in an \ac{RF} environment. The most common sources for \ac{MNO} consist in \ac{RF} interferers that are used to inhibit communications in a security context like prisons or banks to avoid criminal operations. Other common sources of interference are the non calibrated cell sectors from other \acp{MNO} and possible \ac{DoS} attackers. A deep study on interference can be found in \cite{int_stu}.

The analysis of \ac{RF} network performance metrics allows engineers to determine the presence of anomalies. Each \ac{BTS} can work as a sensor where the different metrics, e.g. \ac{RSSI}, will show a modification on its performance under the presence of interference.


As it can be seen in Figure \ref{img:ext_int}, a cluster of cells is depicted. A source of interference is shown, which is transmitting in the same range of frequency as the serving cell. Its interferer coverage can be seen in red in (a) and the degraded sectors are depicted with white and blue stripes. In (b), highlighted with different intensity colors, it is shown the different levels of interference affecting the surrounding cells reduced with increasing distance.




\section{Leveraging Digital Twin Architecture for Network Planning}

In this section, we aim to present an architecture that allows to connect between the real world and simulations. This with the objective of create a network planning framework for 5G an beyond networks.
As it can be seen in Figure \ref{img:architecture}, three layers have been proposed where the lower one corresponds to the \ac{UT}, the second one to the Network layer and the third to the Management layer.


\subsection{Real versus Simulated \ac{RF} data sources}
\label{sec:exp_dig_tw}
According to the review in \cite{dt6gt2p_dt}, the specific use cases we apply are: Network simulation and planning and Network operation and management. Through the use of DT, we aim to enhance the processes of planning and simulating 5G and 6G networks. This will be achieved by utilizing digital replicas of large-scale city blocks and creating a virtual drive test to maintain system robustness and resilience, ultimately improving customer experience and minimizing churn. Additionally, we gather necessary data for the DT foundations, considering data type, time frame, and frequency, along with the mechanisms and tools for data repositories, retrieval, and management. Our focus is on achieving precise physical modeling.

\subsection{Real-world data sources}
\label{sub_rea_wor}

\begin{figure}
\centering
\includegraphics[width=0.35\textwidth]{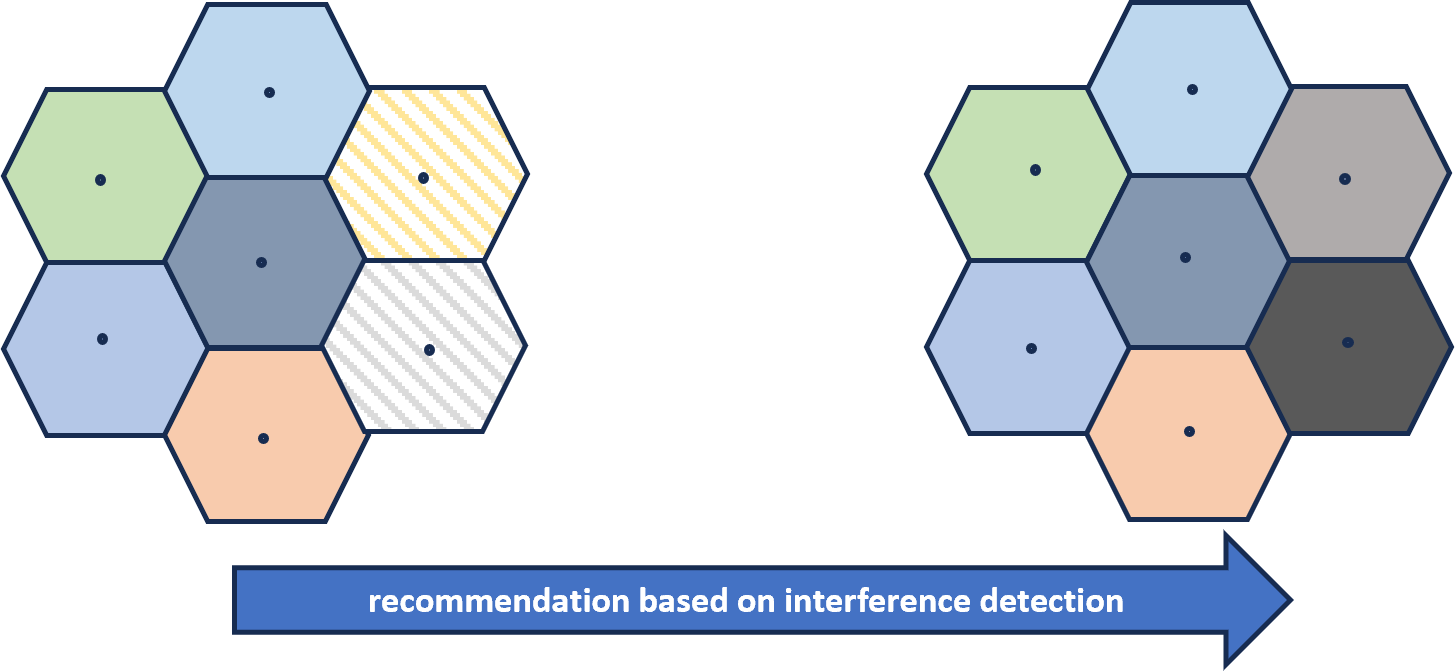}
\caption{Recommending frequency adjustments following interference detection.}\label{img:arc_dig}
\vspace{-2mm}
\end{figure}

As it has been explained in \cite{gra_dig_twi} data is a fundamental part of \ac{DT}. The correct representation of a network in the digital world depends on an adequate data collection of the real-world network. In the case of telecommunication's network, the data sources can come from different sources, like measurement equipment, sensors, and network \ac{KPI}.

\subsubsection{Measurement equipment and \ac{RF} sensors}
Several \ac{RF} measurement equipment, like scanners, can be used for taking data from the wireless environment and this information can be used as a data source for the \ac{DT} model. These values can provide information related to the number of users, data demand, beams features, and can tell us if the \acp{BTS} are closer to our homes and later answer to \ac{EMF} exposure concerns \cite{EMF2015}. 

The \ac{RF} sensors are data sources that can provide information regarding the wireless transmission. These sensors are well specialized equipment used to monitor the network performance, possible sources of interference and other related issues.

\subsubsection{\ac{KPI}}

A network equipment, a \ac{BTS}, a wireless router or even a \ac{UT} in a \ac{MNO} can work as sensor collecting information of the network performance. This information is stored in a \ac{KPI} server known as \ac{OSS}. Several \acp{KPI} can be used for this purpose e.g. the number of \ac{RRC} connections, which give a lot of information of the connection state in a certain period of time. It tells if a specific device has normally finalized the communication or if for some reason, the network could not correctly finish it. 

The \ac{RTWP} serves as a \ac{KPI} providing insights related to the power received from external sources. Additionally, other related metrics such as latency, packet-loss, or throughput provide crucial information about higher-layer network performance \cite{kpi_tes}.

In the proposed architecture in Figure \ref{img:architecture}, all devices in the lower layer, \acp{UT}, \ac{RF} scanners and the measurement equipment, will produce information that can be used in the \ac{DT} implementation. In the network layer, certain intermediate devices can provide information regarding the events and metrics they can access. These devices are \acp{BTS}, satellites, free-cell bridges, \acp{UAV} and others. Depending on the monitoring capabilities the \ac{RIS} devices, these can provide information for \ac{DT}.

\subsection{Simulation world data sources}
\label{sim_wor}
Considering the realm of \ac{RF} network planning, the simulated world can be used as a virtual scenario where several new features can be tested and optimized. Through advanced modelling techniques, \ac{DT} can replicate the complexity of the real-world, where network architects can simulate various deployment strategies to predict potential network issues. In the simulated world, the data sources are essential components that enable the creation of synthetic datasets. These can be generated by robust networking simulators such as NS3, Maltab, NetSim and OMNeT++. Here, researchers can explore diverse scenarios and assess the performance of emerging technologies in controlled environments. 
For this paper, we have selected ATOLL as the simulation tool to evaluate the scenario.
\subsubsection{ATOLL/WINPROP}
ATOLL, a platform by Forsk, is used for the design and optimization of wireless networks supporting a wide range of radio access technologies, including 5G NR, LTE, NB-IoT, UMTS, GSM, CDMA, and incorporating advanced technologies such as MIMO, 3D beamforming, and \ac{mmWave} propagation. WINPROP, a software supported by Altair, is used for planning, DT, and modelling of wireless propagation and mobile networks based on 2D/3D maps in a range of radio access technologies, including 3G/4G/5G networks using mMIMO/beamforming.
These platforms offer a working framework for \ac{MNO} and providers to design, optimize, and plan their networks as technology evolves and user demands expand, enhancing their services \cite{plan_valdemar}.


\subsection{Connections between real-world and simulations}
\label{con_rea_sim}
These are links between the real and simulated planes that allow communication for decision-making and the creation of \ac{DT}. The links can be classified as follows.

\subsubsection{From real to simulated world}
Performance metrics, \acp{KPI} or real-time data can be provided by the deployed equipment and will be a valuable source in simulation environments. Most of \acp{MNO} possesses \acp{OSS} which store all of this information in databases. Additional data as geographical coordinates, real-time localization and radio measurements come from drive or walk tests.

\subsubsection{From simulated to real-world}

\begin{figure}
\centering
\includegraphics[width=0.43\textwidth]{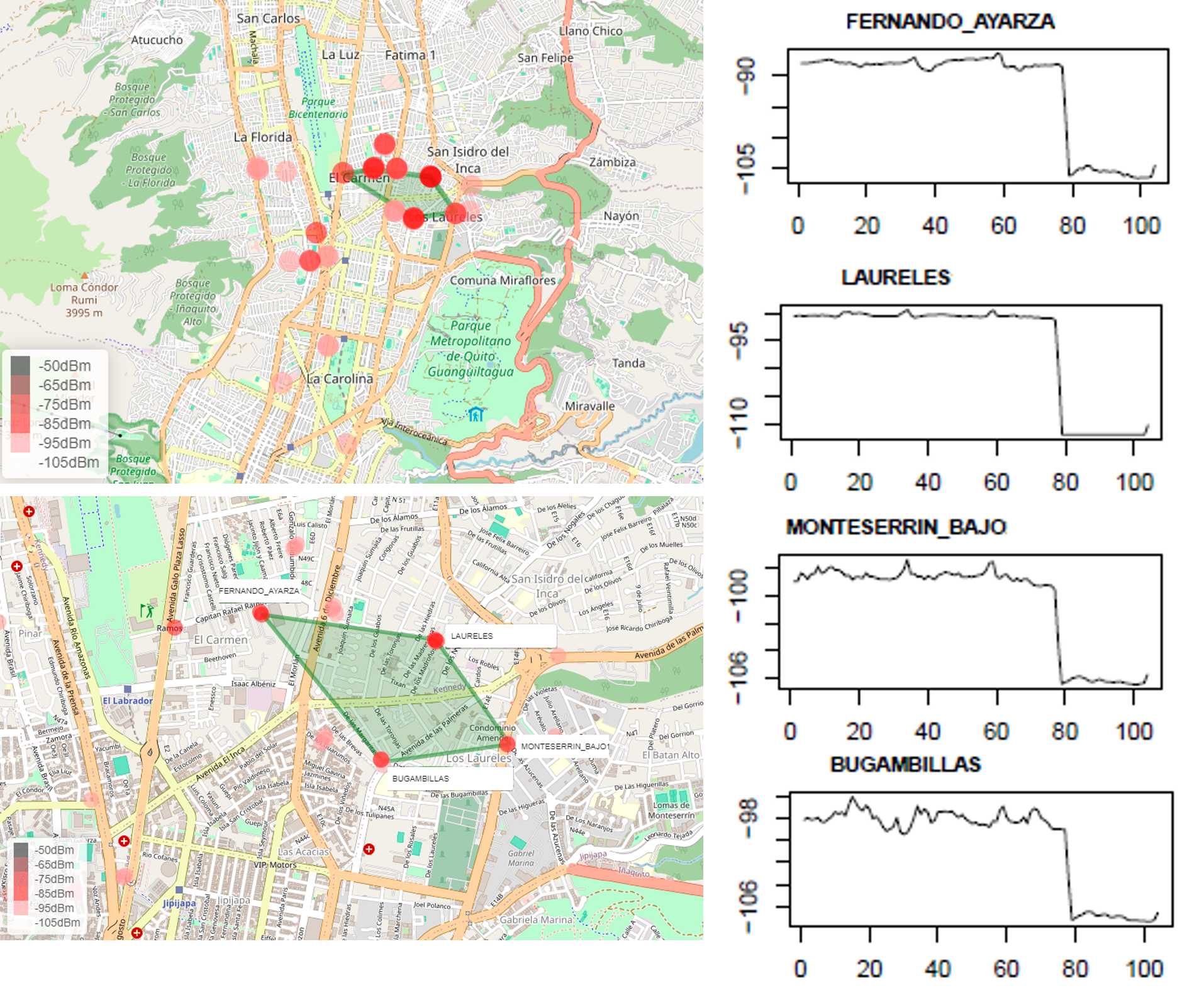}
\vspace{-2mm}
\caption{\ac{RTWP} interference detected in cluster.}\label{img:RTWP1}
\vspace{-3mm}
\end{figure}

A solution that has been tested in simulation can be translated to the real-world, e.g., a new planning scenario that changed because of technical issues in the network deployment. In Figure \ref{img:arc_dig}, it is possible to see how the recommendation would be applied from a simulated scenario to the real world, eliminating the interference from the affected cells.

In Figure \ref{img:architecture}, the connections and interfaces between the different layers, the physical world and \ac{DT} are depicted.

\section{Network planning}

In this section, we provide a framework that incorporates \ac{DT} into network planning as an application of the architecture proposed in Figure \ref{img:architecture}. First, a real-world scenario of data-aided interference identification is explained. Later, this is connected to a simulated scenario using the available interfaces. Finally, the scenario is connected to the management layer.

\subsection{Real-World Interference: A Practical Scenario}

\begin{figure}
\centering
\includegraphics[width=0.4\textwidth]{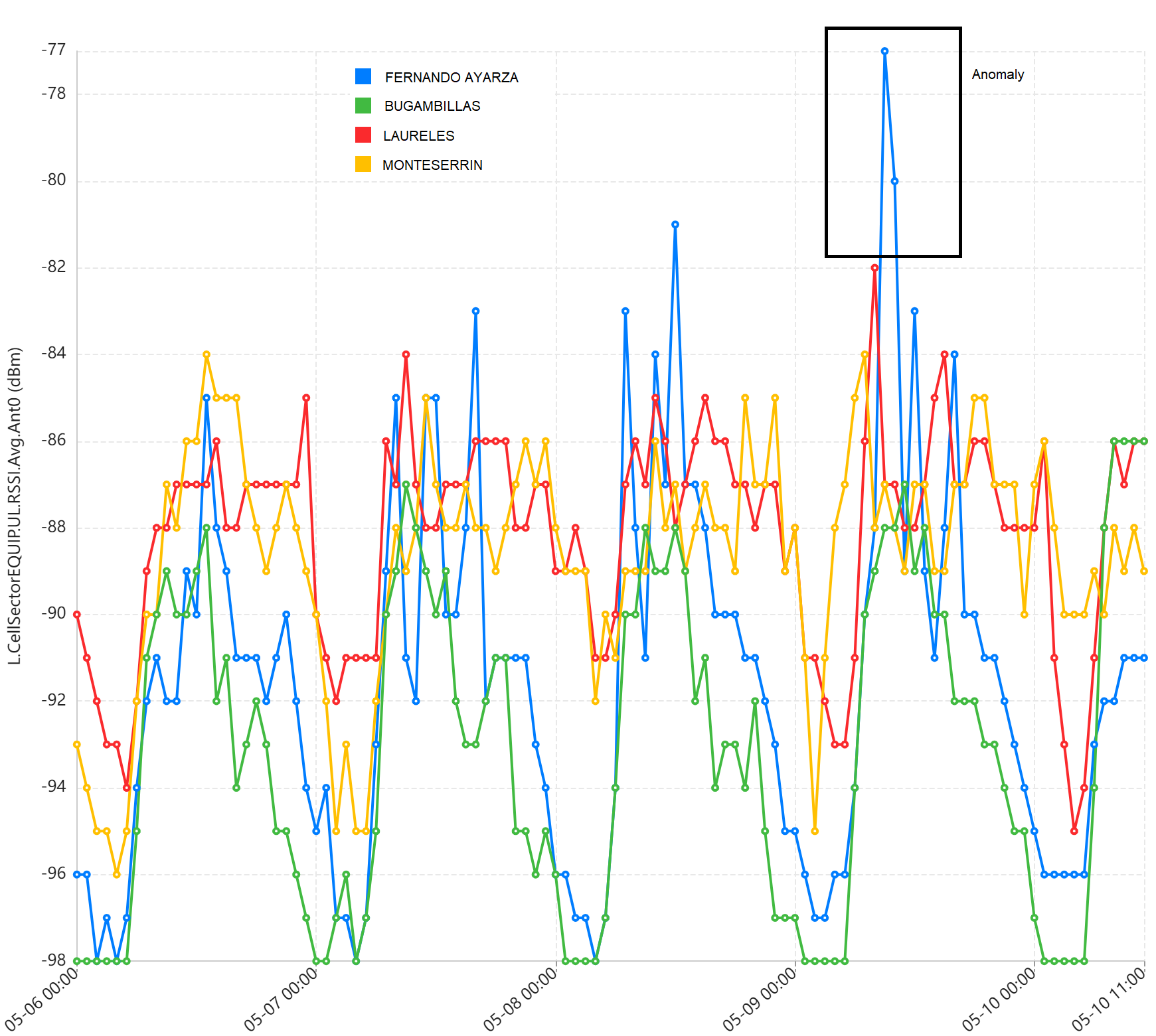}
\caption{\ac{RSSI} interference levels detected in cluster.}\label{img:RSSI1}
\vspace{-2mm}
\end{figure}

As illustrated in Figure \ref{img:RTWP1} and in Figure \ref{img:RSSI1}, in this scenario, a \ac{MNO} detects interference, inferred from anomalies observed in the received power temporal metrics. Given the deployment of several \acp{BTS} throughout the city, measurements of received power levels from various geographic locations are possible. 


From these measurements, the following observations can be made:

\begin{itemize}
  \item Nearest \ac{BTS} to the interferer will have the highest \ac{RTWP} or \ac{RSSI} depending on the technology.
  \item The interference generated by the interferer will produce a similar or correlated effect in the temporal measurement of the received power in their antennas.
  \item The effect of the interference will be reduced with greater distances.
\end{itemize}

Based on these observations, the data generated by this monitoring can be leveraged to train a \ac{ML} classification algorithm. This algorithm has to be capable of identifying anomalies indicative of interference within the network. Once these anomalies are accurately recognized, the objective will be to pinpoint the source of the interference. For this analysis, we have used the clustering algorithm K-means, which is oftentimes used to identify patterns and correlations among the interference measurements obtained by each sensor \ac{BTS}. This algorithm systematically groups similar patterns to accurately identify the sources of interference. The process is explained as follows. The process starts with the data collection of the network performance data coming from the \ac{OSS} where \ac{RTWP}/\ac{RSSI}, a received power metric, are stored with geographical information and the period over which it is gathered. Following the data collection process, the data is cleaned and normalized. For the data analysis, the K-means algorithm groups data points based on similarity, and in this particular case, it distinguishes between different sources based on the received signal associated metrics. Finally, the localization results are validated through cross-referencing with known interference and on-site incidents.

\vspace{-2mm}
\subsection{Coverage area planning}

As it can be seen in the proposed management layer in Figure \ref{img:architecture}, the coverage planning includes the calculation of link budget, path losses, throughput, attenuation and losses between the \ac{UT} and next-generation nodes. These calculations are subject to customer requirements, spectrum availability, specific coverage restrictions, and \ac{EMF} regulatory limits.
It is possible compute the link budget and status of signal in the wireless channel and show the cell ranges for a coverage radio from one \ac{BTS} \cite{plan_valdemar2}. 
Next, we identify the path loss values using \ac{UMa}/\ac{UMi} propagation models recommended at 3GPP 38.901; the cell radius value is estimated as the maximum distance between gnodeB and \ac{UT}. The calculated cell radius determines the number of \ac{BTS} needed in the target area. With the link budget, the next step is determining the locations of each site for the 5G/B5G NR network planning that can be in-depth verified for interference computation in the dense urban/urban area of the zone \cite{plan_valdemar2}. Finally, the management layer includes three entities module, control, \ac{ML} and anomaly detection, and optimization.

\vspace{-1mm}
\subsection{Management layer operation}

In the management layer, the parameters calculations from the planning entity are done using ATOLL/WINPROP. The \ac{ML} block analyzes the data provided by the \ac{OSS} for anomaly detection and prediction, e.g. interference detection. The control entity, through its connections with the network layer, will change the configuration files associated with the recommendations provided by the \ac{ML} entity, e.g. frequency changes, power, codes.

\vspace{-1mm}
\subsection{Extension to a 6G scenario}
\label{subsec:6g_con}

\begin{figure}
\centering
\includegraphics[width=0.45\textwidth]{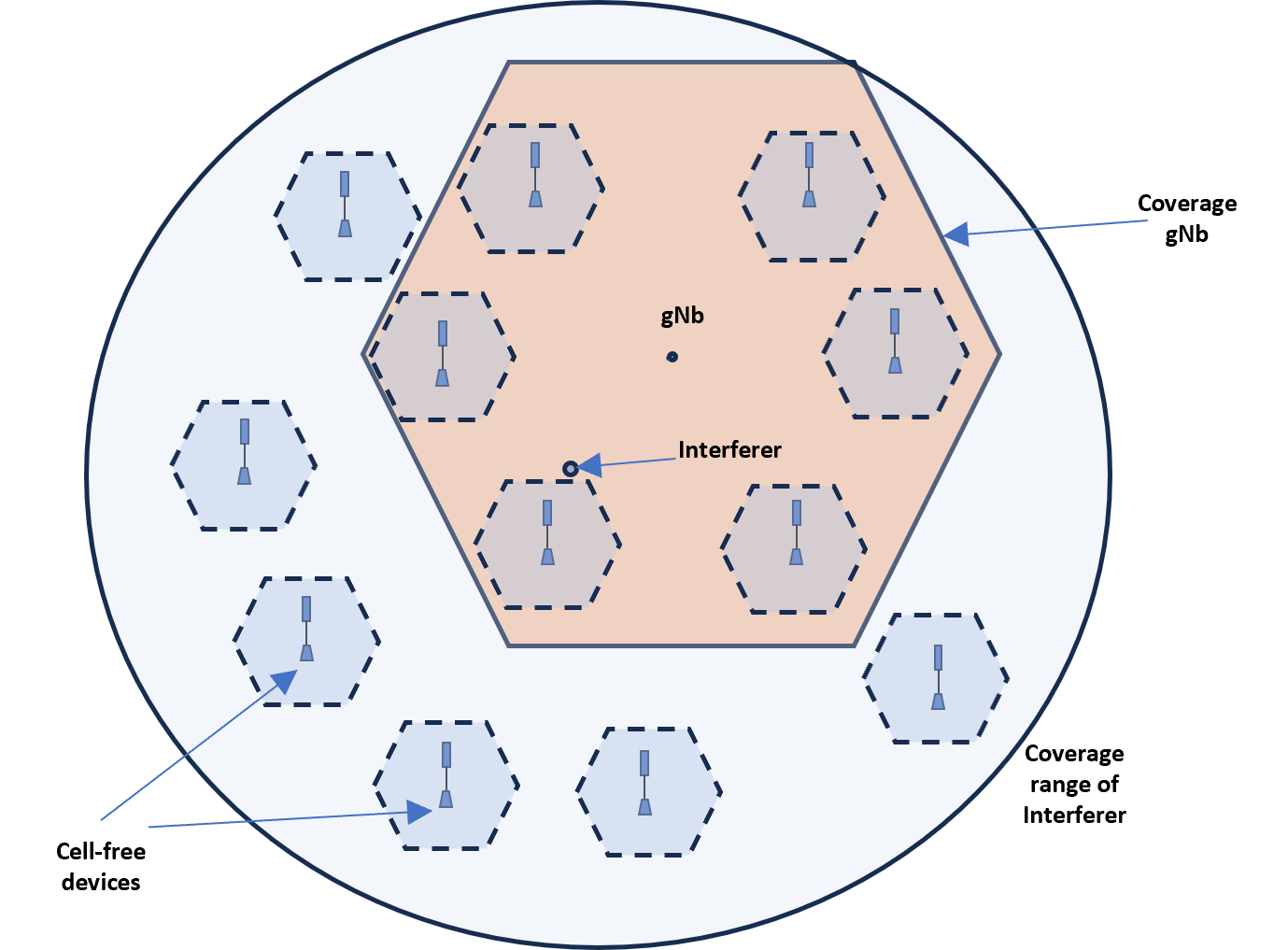}
\caption{Coverage degradation caused by an interference source.}\label{img:cov_deg}
\vspace{-3mm}
\end{figure}

To extend this work to a 6G scenario, it is important to highlight the introduction of new \ac{RF} outstanding new technologies. Among these, cell-free environment emerges as an important change, where communications rely solely on user devices and passive antennas, eliminating traditional cell-based infrastructure. The introduction of heterogeneous network elements and terminals promises to increase the data set for analysis. These devices can provide valuable information about network conditions, including more details about interference and localization, as can be seen in Figure \ref{img:cov_deg}. With an expanded network, the data increases significantly, enhancing the precision of the methods. Increasing the number of antennas within the affected area by interference will enhance the effectiveness of detection algorithms.

\vspace{-1mm}
\section{Results}
\label{sec:results}

\begin{figure}
\centering
\includegraphics[width=0.45\textwidth]{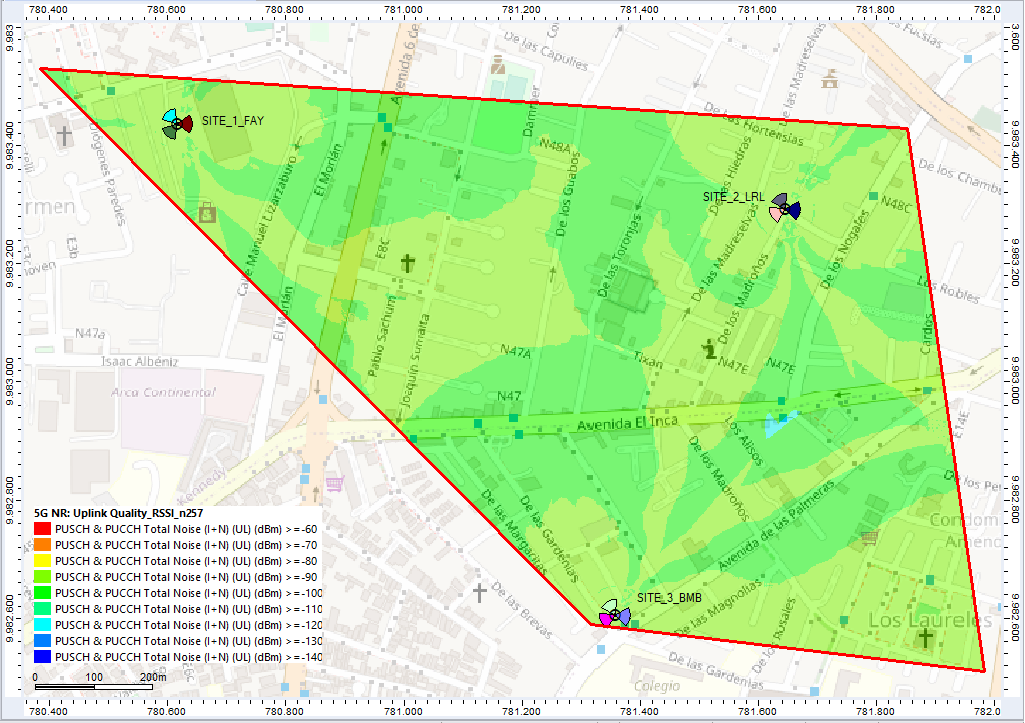}
\caption{\ac{RSSI} Simulation at 28 GHz (ATOLL).}\label{img:RSSI28G}
\vspace{-3mm}
\end{figure}

\begin{figure}
\centering
\includegraphics[width=0.43\textwidth]{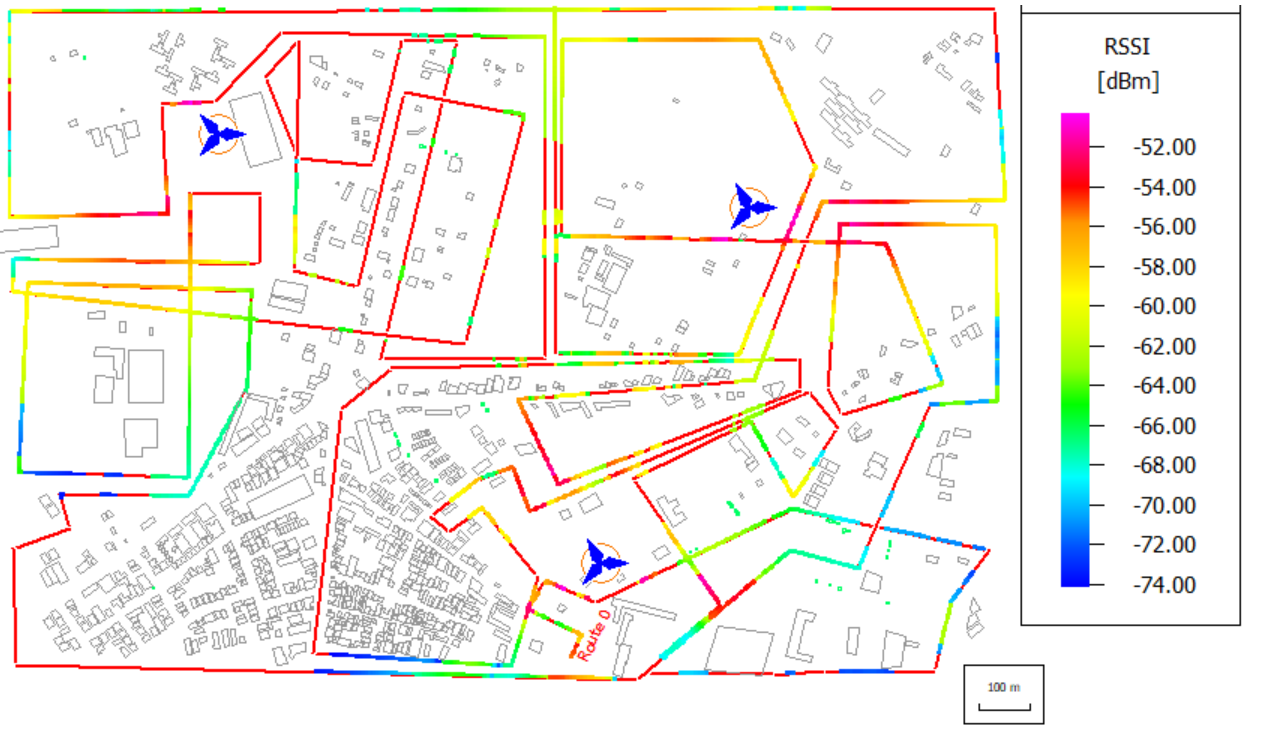}
\vspace{-5mm}
\caption{\ac{RSSI} Physical Trajectory (virtual drive test) at 28 GHz (WINPROP).}\label{img:RSSI28G_2}
\end{figure}

\begin{figure}
\centering
\includegraphics[width=0.43\textwidth]{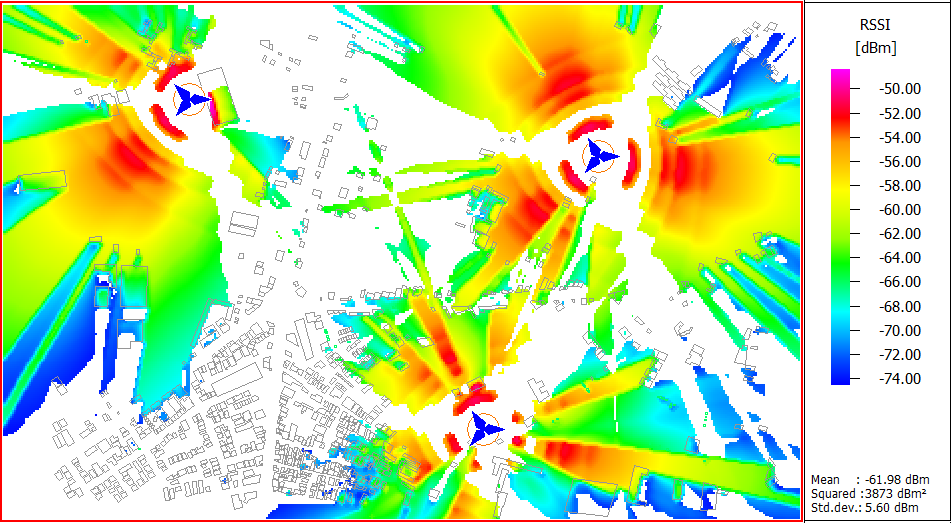}
\vspace{-2mm}
\caption{\ac{RSSI} \ac{DT} at 28 GHz (WINPROP).}\label{img:RSSI28G_3}
\vspace{-4mm}
\end{figure}

As illustrated in Figure \ref{img:architecture}, the scenario under examination is highlighted with dots, focusing exclusively on data derived from the scenario depicted as interference localization. In this scenario, the intermediate devices are \acp{BTS}, and the data considers statistics from regular phone users. 
It can be noted that having the fingerprints of real samples, whether from \acp{KPI} or quality parameters, coupled with information from simulation generated from physical data, resulted in an average outcome in \ac{DT} very close to reality for each metric, such as \ac{RSSI}, \ac{RTWP}, \ac{SINR}, and throughput. Likewise, it is observed that coherence and propagation logic are preserved, as increasing the transmission frequency value leads to a smaller coverage range, increased bandwidth and capacity, resulting in an enhanced user service experience. Finally, by increasing precision with DT results, the detection level of network anomalies and distortions will be improved, geolocating them and measuring their value within the range of their thresholds, providing excellent optimization for our management and quality processes.
The results obtained from the \ac{DT} process can be seen in Table \ref{NombreTabla222}, where the calculated mean for the DT results in comparison to the values from simulations showed high accuracy across all metrics, particularly in \ac{RTWP}, \ac{SINR}, and throughput. Additionally, at the frequency band level, higher accuracy was achieved for the \ac{mmWave} band (28 GHz) across all metrics. Regarding the \ac{RSSI} plots, Figure \ref{img:RSSI28G} illustrates that the simulation provides a good accuracy of interference behavior in the cluster. Furthermore, Figure \ref{img:RSSI28G_2} provides more precise virtual path trajectories, considering the Ray Tracing algorithm and 3D maps of buildings and obstacles. Finally, Figure \ref{img:RSSI28G_3} presents the results of the \ac{DT} process, wherein better accuracy is achieved compared to simulation and virtual paths, yielding values closer to the actual \ac{RSSI} behavior in a 5G/B5G network. 

\begin{table*}[ht]
\footnotesize
\caption{Simulation results for the \ac{DT} process.} 
\label{NombreTabla222}
\centering
\begin{tabular}{|l|c|cc|cc|cc|}
\hline
\multicolumn{1}{|c|}{\multirow{-2}{*}{\textbf{S.No}}}&
  \multicolumn{1}{c|}{\textbf{\begin{tabular}[c]{@{}c@{}}Parameters \\ with Units\end{tabular}}} &
  \multicolumn{1}{c|}{\textbf{\begin{tabular}[c]{@{}c@{}}Simul. results\\ for 3.5 GHz\end{tabular}}} &
  \textbf{\begin{tabular}[c]{@{}c@{}}Simul. results\\ for 28 GHz\end{tabular}} &
  \multicolumn{1}{c|}{\textbf{\begin{tabular}[c]{@{}c@{}}Exp. values\\ for 3.5 GHz\end{tabular}}} &
  \textbf{\begin{tabular}[c]{@{}c@{}}Exp. values\\ for 28 GHz\end{tabular}} &
  \multicolumn{1}{c|}{\textbf{\begin{tabular}[c]{@{}c@{}}Exp. values\\ for 3.5 GHz\end{tabular}}} &
  \textbf{\begin{tabular}[c]{@{}c@{}}Exp. values\\ for 28 GHz\end{tabular}} 
  \\
  \hline
$1$ &
  \multicolumn{1}{c|}{RSSI (dBm)} &
  \multicolumn{1}{c|}{-89.06} &
  $-90.65$ &
  \multicolumn{1}{c|}{-80} &
  \multicolumn{1}{c|}{-85} &
  \multicolumn{1}{c|}{-99.66} &
  \multicolumn{1}{c|}{-92.97} 
  \\ \hline
$2$ &
  \multicolumn{1}{c|}{RTWP (dBm)} &
  \multicolumn{1}{c|}{-102} &
  $-102$ &
  \multicolumn{1}{c|}{-82} &
  \multicolumn{1}{c|}{-98}  &
  \multicolumn{1}{c|}{-103} &
  \multicolumn{1}{c|}{-103} 
  \\ \hline
$3$ &
  {SINR} (dB) &
  \multicolumn{1}{c|}{9.05} &
  $-0.57$ &
  \multicolumn{1}{c|}{10} &
  \multicolumn{1}{c|}{11} &
  \multicolumn{1}{c|}{9.60} &
  \multicolumn{1}{c|}{9.08} 
  \\ \hline
$4$ &
  DL:Throughput   (Mbits/s) &
  \multicolumn{1}{c|}{500} &
  $1000$ &
  \multicolumn{1}{c|}{622.7} &
  \multicolumn{1}{c|}{1245} &
  \multicolumn{1}{c|}{622.7} &
  \multicolumn{1}{c|}{1245} 
  \\ \hline
$5$ &
  UL:Throughput   (Mbits$/$s) &
  \multicolumn{1}{c|}{500} &
  $1000$ &
  \multicolumn{1}{c|}{622.7} &
  \multicolumn{1}{c|}{1245} &
  \multicolumn{1}{c|}{622.7} &
  \multicolumn{1}{c|}{1245} 
  \\ \hline 
\end{tabular}
\end{table*}

%

%
\acresetall
\vspace{-4mm}
\section{Conclusions}
\label{sec:conclusions}
\vspace{-2mm}
In this paper, we have highlighted the importance of addressing the challenges of interference problems for next generation wireless networks. The integration of digital twins provides perspectives on the performance of interference sources, enabling MNOs to anticipate network performance and predict the impact of external anomalies e.g. interference. By detecting impairments and comparing them across sectors and network elements, service providers can distinguish normal behaviour from degradation, facilitating mitigation operation. Finally, exploring the connections between real data and simulations presents new opportunities to improve network planning and self-management.

%
\vspace{-3mm}
\section*{Acknowledgment}
{\small{Supported by the IN2CCAM project funded by the European Union’s Horizon Europe research and innovation program under grant agreement No 101076791. The authors acknowledge the use of the planning tool ATOLL from FORSK and WINPROP from ALTAIR which played a pivotal role in the analysis phase of this investigation.}}
\vspace{-3mm}
\printbibliography

@ARTICLE{int_stu,
  author={Siddiqui, Maraj Uddin Ahmed and Qamar, Faizan and Ahmed, Faisal and Nguyen, Quang Ngoc and Hassan, Rosilah},
  journal={IEEE Access}, 
  title={Interference Management in 5G and Beyond Network: Requirements, Challenges and Future Directions}, 
  year={2021},
  volume={9},
  number={},
  pages={68932-68965},
  doi={10.1109/ACCESS.2021.3073543}}

@ARTICLE{holo_antena,
  author={An, Jiancheng and Yuen, Chau and Huang, Chongwen and Debbah, Mérouane and Poor, H. Vincent and Hanzo, Lajos},
  journal={IEEE Commun. Lett.}, 
  title={A Tutorial on Holographic MIMO Communications—Part II: Performance Analysis and Holographic Beamforming}, 
  year={2023},
  volume={27},
  number={7},
  pages={1669-1673},
  doi={10.1109/LCOMM.2023.3278682}}

@ARTICLE{fuid_antenna,
  author={Wong, Kai-Kit and Shojaeifard, Arman and Tong, Kin-Fai and Zhang, Yangyang},
  journal={IEEE Trans. on Wireless Commun.}, 
  title={Fluid Antenna Systems}, 
  year={2021},
  volume={20},
  number={3},
  pages={1950-1962},
  doi={10.1109/TWC.2020.3037595}}

@INPROCEEDINGS{VLC,
  author={Dwivedy, Prashant and Dixit, Vipul and Kumar, Atul},
  booktitle={2023 Int. Conf. on Computer, Electronics \& Electrical Eng. \& their Applications (IC2E3)}, 
  title={A Survey on Visible Light Communication for 6G: Architecture, Application and Challenges}, 
  year={2023},
  volume={},
  number={},
  pages={1-6},
  doi={10.1109/IC2E357697.2023.10262462}}

@article{HUANG2023106,
title = {From Terahertz Imaging to Terahertz Wireless Communications},
journal = {Engineering},
volume = {22},
pages = {106-124},
year = {2023},
issn = {2095-8099},
doi = {https://doi.org/10.1016/j.eng.2022.06.023},
url = {https://www.sciencedirect.com/science/article/pii/S2095809922006361},
author = {Yi Huang and Yaochun Shen and Jiayou Wang},
}

@article{sandouno2023novel,
    title={{A novel approach for ray tracing optimization in wireless communication}},
    author={Sandouno, Bernard Tamba and Alsaba, Yamen and Barakat, Chadi and Dabbous, Walid and Turletti, Thierry},
    journal={Computer Commun.},
    volume={209},
    pages={309--319},
    year={2023},
    doi={10.1016/j.comcom.2023.07.016},
    publisher={Elsevier}
}

@inproceedings{seidel1992ray,
    title={{A ray tracing technique to predict path loss and delay spread inside buildings}},
    author={Seidel, Scott Y and Rappaport, Theodore S},
    booktitle={[Conference Record] GLOBECOM'92-Communications for Global Users: IEEE},
    pages={649--653},
    year={1992},
    doi={10.1109/GLOCOM.1992.276436},
    organization={IEEE}
}

@techreport{3GPPTR38901,
    author = {{European Telecommunications Standards Institute}},
    title = {{5G; Study on channel model for frequencies from 0.5 to 100 GHz (3GPP TR 38.901 version 14.1.1 Release 14)}},
    institution = {ETSI},
    month = {8},
    number = {138 901 V14.1.1},
    type = {TR},
    year = {2017},
}

@ARTICLE{dt_net1,
  author={Lin, Xingqin and Kundu, Lopamudra and Dick, Chris and Obiodu, Emeka and Mostak, Todd and Flaxman, Mike},
  journal={IEEE Commun. Magazine}, 
  title={6G Digital Twin Networks: From Theory to Practice}, 
  year={2023},
  volume={61},
  number={11},
  pages={72-78},
  doi={10.1109/MCOM.001.2200830}}

@INPROCEEDINGS{int_2023,
  author={Rubab, Naila and Zeb, Shah and Mahmood, Aamir and Hassan, Syed Ali and Ashraf, Muhammad Ikram and Gidlund, Mikael},
  booktitle={2022 IEEE 12th Sensor Array and Multichannel Signal Processing Workshop (SAM)}, 
  title={Interference Mitigation in RIS-assisted 6G Systems for Indoor Industrial IoT Networks}, 
  year={2022},
  volume={},
  number={},
  pages={211-215},
  doi={10.1109/SAM53842.2022.9827809}}

@INPROCEEDINGS{con_2020,
  author={Moysen, Jessica and Ahmed, Furqan and García-Lozano, Mario and Niëmela, Jarno},
  booktitle={2020 European Conf. on Networks and Commun. (EuCNC)}, 
  title={Unsupervised learning for detection of mobility related anomalies in commercial LTE networks}, 
  year={2020},
  volume={},
  number={},
  pages={111-115},
  doi={10.1109/EuCNC48522.2020.9200970}}

@ARTICLE{6G_ML,
  author={Noman, Hafiz Muhammad Fahad and Hanafi, Effariza and Noordin, Kamarul Ariffin and Dimyati, Kaharudin and Hindia, Mhd Nour and Abdrabou, Atef and Qamar, Faizan},
  journal={IEEE Access}, 
  title={Machine Learning Empowered Emerging Wireless Networks in 6G: Recent Advancements, Challenges and Future Trends}, 
  year={2023},
  volume={11},
  number={},
  pages={83017-83051},
  doi={10.1109/ACCESS.2023.3302250}}

@INPROCEEDINGS{cha_mod_2021,
  author={Walidainy, Hubbul and Adriman, Ramzi and Away, Yuwaldi and Nasaruddin, Nasaruddin},
  booktitle={2021 Int. Conf. on Computer System, Information Tech., and Elect. Eng. (COSITE)}, 
  title={Channel Modeling for 6G Communications: A Survey}, 
  year={2021},
  volume={},
  number={},
  pages={1-6},
  doi={10.1109/COSITE52651.2021.9649545}}

@ARTICLE{plan_valdemar2,
  author={Farré, Valdemar and Vega, Jose and Carvajal, Henry},
  journal={Eng. Proceedings-MDPI}, 
  title={5G NR Radio Network Planning at 3.5 GHz and 28 GHz in a Business/Dense Urban Area from the North Zone in Quito City}, 
  year={2023},
  volume={47},
  number={24},
  pages={},
  doi={10.3390/engproc2023047024}}

@ARTICLE{plan_valdemar,
  author={Farré, Valdemar and Vega, José and Paredes, Cecilia and Grijalva, Felipe and Moya, Diana},
  journal={IEEE Access}, 
  title={Comparative Evaluation of Radio Network Planning for Different 5G-NR Channel Models on Urban Macro Environments in Quito City}, 
  year={2024},
  volume={12},
  number={},
  pages={5708-5730},
  doi={10.1109/ACCESS.2024.3350182}}

@ARTICLE{EMF2015,
  author={Wali, Sangin Qahtan and Sali, Aduwati and Allami, Jaafar K. and Osman, Anwar Faizd},
  journal={IEEE Access}, 
  title={RF-EMF Exposure Measurement for 5G Over Mm-Wave Base Station With MIMO Antenna}, 
  year={2022},
  volume={10},
  number={},
  pages={9048-9058},
  doi={10.1109/ACCESS.2022.3143805}}

@INPROCEEDINGS{gra_dig_twi,
  author={Zhu, Yanhong and Chen, Danyang and Zhou, Cheng and Lu, Lu and Duan, Xiaodong},
  booktitle={2021 IEEE 1st International Conference on Digital Twins and Parallel Intelligence (DTPI)}, 
  title={A knowledge graph based construction method for Digital Twin Network}, 
  year={2021},
  volume={},
  number={},
  pages={362-365},
  doi={10.1109/DTPI52967.2021.9540177}}

@INPROCEEDINGS{kpi_tes,
  author={Saha, Niloy and James, Alexander and Shahriar, Nashid and Boutaba, Raouf and Saleh, Aladdin},
  booktitle={NOMS 2022-2022 IEEE/IFIP Network Operations and Management Symposium}, 
  title={Demonstrating Network Slice KPI Monitoring in a 5G Testbed}, 
  year={2022},
  volume={},
  number={},
  pages={1-3},
  doi={10.1109/NOMS54207.2022.9789904}}

@ARTICLE{emp6g_dt,
  author={Kuruvatti, Nandish P. and Habibi, Mohammad Asif and Partani, Sanket and Han, Bin and Fellan, Amina and Schotten, Hans D.},
  journal={IEEE Access}, 
  title={Empowering 6G Communication Systems With Digital Twin Technology: A Comprehensive Survey}, 
  year={2022},
  volume={10},
  number={},
  pages={112158-112186},
  doi={10.1109/ACCESS.2022.3215493}}

@ARTICLE{dt6gt2p_dt,
  author={Lin, Xingqin and Kundu, Lopamudra and Dick, Chris and Obiodu, Emeka and Mostak, Todd and Flaxman, Mike},
  journal={IEEE Commun. Magazine}, 
  title={6G Digital Twin Networks: From Theory to Practice}, 
  year={2023},
  volume={61},
  number={11},
  pages={72-78},
  doi={10.1109/MCOM.001.2200830}}
\end{document}